\RequirePackage{fix-cm}
\documentclass[11pt]{article}

\usepackage{fullpage}
\usepackage[small,bf]{caption}
\usepackage{amsfonts,dsfont,amssymb,amsbsy,amsmath,paralist,theorem,bm,ifthen,color}
\usepackage[pdfstartview=FitH,bookmarksnumbered,unicode,bookmarksopen=true]{hyperref}
\usepackage{graphicx}
\usepackage{epstopdf}
\usepackage{multirow}
\usepackage{subfigure,relsize}
\usepackage{mathrsfs}
\usepackage{bm}
\usepackage{stfloats}
\usepackage{appendix}
\usepackage{url}
\usepackage[noadjust]{cite}
\usepackage{cases,booktabs}
\usepackage{graphicx,algorithmic,algorithm,relsize}
\usepackage[justification=centering]{caption} 
\usepackage[table]{xcolor}

\newtheorem{lem}{Lemma}
\newtheorem{prop}{Proposition}

\newtheorem{rem}{Remark}

\title{Waveguide Deployment Optimization for Pinching-Antenna Systems}

\author{Yanqing Xu, 
        Zhiguo Ding, 
        Xiu Yin Zhang, 
        and Tsung-Hui Chang
        \thanks{\smaller[1] Y. Xu is with the School of Electronic and Information Engineering, South China University of Technology, Guangzhou 510641, China (email: xuyanqing@scut.edu.cn).}
        \thanks{\smaller[1] Z. Ding is with the School of Electrical \& Electronic Engineering, Nanyang Technological University, 639798, Singapore  (e-mail: zhiguo.ding@ntu.edu.sg).}
        \thanks{\smaller[1] X. Y. Zhang is with the School of Microelectronics, South China University of Technology, Guangzhou 510000, China, and also with the Pazhou Laboratory, Guangzhou 510330, China (e-mail: eexyz@scut.edu.cn).}
        \thanks{\smaller[1] T.-H. Chang is with the School of Artificial Intelligence, The Chinese University of Hong Kong, Shenzhen 518172, China (email: changtsunghui@cuhk.edu.cn).}
    	\thanks{\smaller[1] This paper has been submitted to the IEEE for possible publication. Copyright may be transferred without notice, after which this version may no longer be accessible.}}

\begin{document}
\maketitle

\begin{abstract}
Most existing studies on pinching-antenna systems optimize the
antenna positions or activation states under a prescribed waveguide
deployment, where mutually parallel waveguides are commonly adopted.
However, it remains unclear whether, and under what conditions, such a
deployment is optimal from a communication-performance perspective. To
fill this gap, this paper investigates long-term waveguide deployment
optimization for pinching-antenna systems over a rectangular service
region. We first consider the fundamental two-waveguide case and derive
a closed-form sufficient condition, jointly determined by the
waveguide height and the service-region aspect ratio, under which a
deployment parallel to the longer side of the service region is
globally optimal. We then consider the general multi-waveguide case,
where the orientation and offset of each waveguide are jointly optimized
without prescribing any deployment pattern, and develop an adaptive
exchange algorithm for the resulting nonconvex optimization problem. We
further show that, within the parallel class, the waveguides should be
aligned with the longer side of the service region, while in the
large-height regime, the globally optimal unrestricted deployment
asymptotically concentrates all waveguides on the centerline along the
longer side. Numerical results validate the proposed algorithm and
reveal how the preferred deployment geometry depends on the waveguide
height and service-region shape. In particular, nonparallel deployments
can provide noticeable gains in intermediate-height regimes, whereas
parallel or center-concentrated deployments become favorable in other
regimes.
\end{abstract}


\section{Introduction}
\label{sec:introduction}

Pinching-antenna systems have recently emerged as a
promising flexible-antenna architecture for next-generation wireless
communications \cite{suzuki2022pinching,ding2024flexible}. By activating dielectric perturbations
along a low-loss waveguide, radiating points, referred to as pinching
antennas (PAs), can be created and repositioned over a large spatial
range. This capability allows the effective radio-access points to be
adapted to the locations of the served users and provides a new spatial
degree of freedom for communication-system design.

Existing studies have primarily exploited this flexibility by optimizing
the PA positions or activation states over prescribed waveguides \cite{yang2025pinching, xu2026generalized,lu2026survey}. 
For single-waveguide systems, PA-position optimization has been investigated
for rate maximization \cite{xu2025rate,che2025throughput}, multi-user transmission \cite{wang2025antenna,xu2025qos}, and practical propagation
effects such as in-waveguide attenuation
\cite{xu2026pinching-attenuation,tyrovolas2025performance}. Related analytical
studies have further characterized the optimal PA locations under
representative transmission strategies \cite{ding2026analytical}. These
works have demonstrated the performance benefits of adapting the
radiation locations along a waveguide, but the supporting waveguide
geometry itself is generally treated as given.
PA-position design has also been extended to multi-waveguide systems,
with recent studies jointly considering PA activation, pinching beamforming,
transmit beamforming, and conventional resource-allocation variables such
as power allocation
\cite{Wang2026AntennaActivation,Zhao2026WaveguideDivision,
Shan2026Multigroup,Ren2026MultipleAccess,qian2026pinching,gan2026dualpa}.
In these systems, mutually parallel or regularly arranged waveguides are
commonly adopted because of their simple installation and regular
geometry. Nevertheless, the waveguide locations and orientations are
still typically specified in advance. Consequently, these studies mainly
address how the PAs should be configured over a given multi-waveguide
architecture, rather than how the waveguides themselves should be
deployed.

Beyond link-level PA configuration, recent studies have begun to exploit PA spatial reconfigurability from a broader network perspective. For example, PA-assisted environment-division multiple access (EDMA) uses controllable radiation locations to shape desired and interference links for multi-user transmission \cite{ding2026environment}, while multi-cell PA transmission exploits such spatial flexibility to reduce the transmit power required for inter-cell interference management \cite{ding2026toward}. From a more general network-design perspective, \cite{xu2026pinching-scis} further identified the spatial deployment of waveguides, including their positions, orientations, lengths, and heights, as an important degree of freedom. These developments motivate extending the design space from PA positioning along prescribed waveguides to the geometry of the waveguides themselves. Initial studies have begun to explore this direction. In \cite{ding2026pinching}, movable waveguides were introduced, with the waveguide locations, PA positions, and transmit beamforming jointly optimized to provide additional flexibility in shaping the overall array geometry. Complementarily, \cite{lu2025dual} investigated the impact of waveguide orientation by comparing parallel and orthogonal dual-waveguide deployments for physical-layer security.

Despite these advances, the long-term deployment of waveguides remains largely unexplored. In particular, although mutually parallel waveguides are widely adopted in multi-waveguide systems, it remains unclear whether, and under what conditions, such a deployment is optimal from a communication-performance perspective. This motivates us to treat the waveguide geometry itself as a long-term infrastructure design variable. Specifically, the locations and orientations of the waveguides are determined before the instantaneous service location is known and remain fixed after deployment, while the active PA on each waveguide can be reconfigured on demand according to the service location. Accordingly, we study the deployment of multiple waveguides over a continuous service region and optimize their geometry based on the worst-location communication performance achievable after PA reconfiguration. The resulting deployment problem is challenging because the waveguide locations and orientations must be jointly designed to ensure reliable service across the entire region after accounting for subsequent PA reconfiguration.

\begin{figure*}[t]
    \centering
    \includegraphics[width=0.98\textwidth]{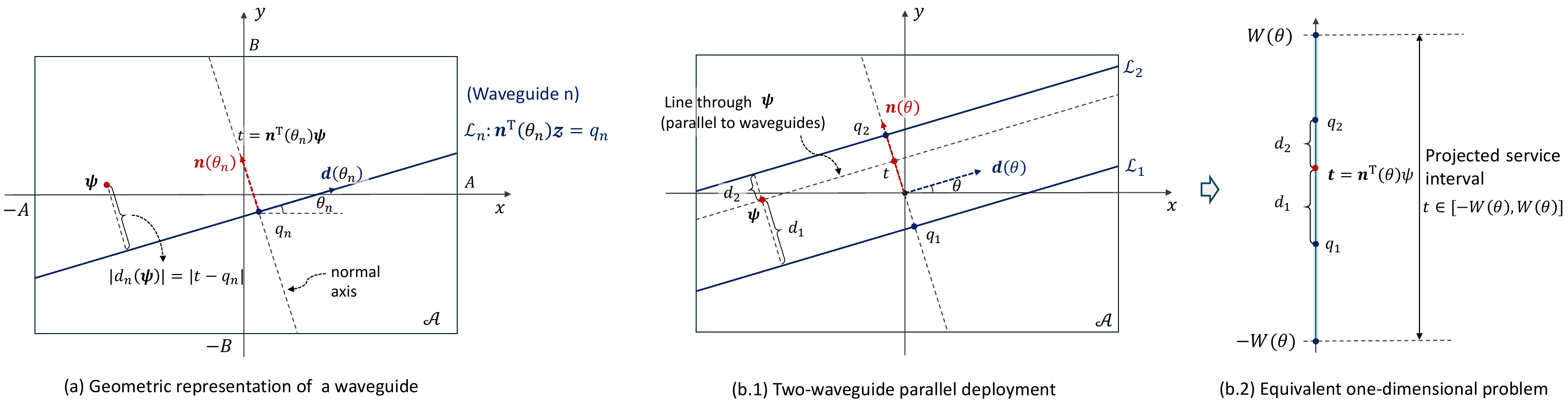}          \captionsetup{justification=justified, singlelinecheck=false, font=small}	      
    \caption{Top-view illustration of the waveguide geometry and its
    parallel two-waveguide specialization. (a) Geometric representation of a waveguide, where
    $\theta_n$ denotes the waveguide orientation, $q_n$ is the signed
    offset along the normal direction, and
    $d_n(\boldsymbol\psi)$ is the signed perpendicular displacement from
    service location $\boldsymbol\psi$ to waveguide $\mathcal L_n$.
    (b.1) Two-waveguide parallel deployment with common orientation
    $\theta$, where $t=\mathbf n^T(\theta)\boldsymbol\psi$ denotes the
    normal coordinate of the service location. (b.2) Equivalent
    one-dimensional representation over the projected service interval
    $[-W(\theta),W(\theta)]$.} \label{fig:parallel_projection}
\end{figure*}

Accordingly, we study the deployment of multiple waveguides over a
continuous service region without assuming a specific spatial user
distribution, and optimize their geometry based on the worst-location
communication performance achievable after PA reconfiguration to
ensure ubiquitous regional coverage.
The main contributions are summarized as follows.
\begin{itemize}
    \item We first focus on the fundamental two-waveguide case and obtain
    analytical deployment results. Within the parallel-waveguide class,
    we characterize the optimal orientation and derive the optimal
    waveguide separation in closed form. We then establish an explicit
    geometry-dependent sufficient condition, jointly determined by the
    waveguide height and service-region aspect ratio, under which the
    resulting parallel deployment is also globally optimal over arbitrary
    waveguide locations and orientations. These results provide a rigorous
    justification for the commonly adopted parallel-waveguide architecture
    and identify a broad parameter regime in which such a deployment is
    globally optimal.

    \item We next consider the general multi-waveguide case, where the waveguide locations and orientations are jointly optimized without prescribing any deployment pattern. We develop an adaptive exchange algorithm to solve the resulting nonconvex problem. We then characterize two important special regimes. Within the parallel-waveguide class, we show that the optimal orientation is along the longer side of the service region. Moreover, in the large-height regime, we prove that the globally optimal unrestricted deployment asymptotically concentrates all waveguides on the centerline along the longer side. These results show that, although nonparallel deployments can be beneficial in general, the widely adopted parallel architecture naturally emerges as optimal under important system regimes.

    \item Finally, numerical results validate the analytical findings and demonstrate the effectiveness of the proposed adaptive exchange algorithm. They reveal how the preferred waveguide geometry varies with the waveguide height and service-region shape. In particular, nonparallel deployments can provide noticeable gains in intermediate-height regimes, whereas parallel or center-concentrated deployments become favorable in other regimes. These results further clarify when different waveguide deployment structures should be adopted in practice.
\end{itemize}

The remainder of this paper is organized as follows. Section \ref{sec:system_model} presents the system model and formulates the waveguide deployment problem. Section \ref{sec:two_waveguide} investigates the two-waveguide case and derives the corresponding analytical results. Section \ref{sec:generalN} considers the general multi-waveguide case and develops the proposed adaptive exchange algorithm together with structural results for important special regimes. Section \ref{sec:numerical} presents the numerical results, and Section \ref{sec:conclusion} concludes the paper.

\section{System Model and Problem Formulation}
\label{sec:system_model}

Consider a rectangular service region on the horizontal plane,
$\mathcal A=[-A,A]\times[-B,B]$, with $A\ge B$, where $D_x=2A$ and $D_y=2B$ denote its side lengths. A baseband processor is connected to $N\ge2$ dielectric waveguides deployed at height $d_v$ above the service plane and coordinates signal transmission and PA activation across the waveguides. The waveguides are regarded as long-term network infrastructure and remain fixed in their horizontal positions and orientations after deployment. During network operation, one PA is activated on each waveguide and its location is reconfigured according to the currently scheduled service location. Thus, waveguide deployment and PA activation operate on different time scales.

We consider straight waveguides that are sufficiently long such that,
for every service location in $\mathcal A$, the orthogonal projection
onto each waveguide lies within the corresponding physical waveguide
segment. Under this assumption, replacing a finite waveguide by its
infinite supporting line does not change the minimum distance from any
service location. Hence, the horizontal projection of waveguide
$n\in\{1,\ldots,N\}$ can be represented by an infinite line
$\mathcal L_n$. For an orientation angle $\theta$, define the unit
vectors along and normal to a waveguide as
\begin{align}
\boldsymbol d(\theta)&=[\cos\theta,\sin\theta]^{\top},\\
\boldsymbol n(\theta)&=[-\sin\theta,\cos\theta]^{\top},
\label{eq:direction_normal}
\end{align}
respectively. As shown in Fig. \ref{fig:parallel_projection}(a), a general waveguide can then be represented as
\begin{align}
\mathcal L_n(\theta_n,q_n)
=\{\boldsymbol z\in\mathbb R^2:
\boldsymbol n^{\top}(\theta_n)\boldsymbol z=q_n\},
\label{eq:general_line}
\end{align}
where $0\le\theta_n\le\pi$ denotes the waveguide orientation and
$q_n$ denotes its signed offset along the normal direction. The
orientations $\theta_n=0$ and $\theta_n=\pi$ are physically
equivalent. For a service location
$\boldsymbol\psi=[x,y]^{\top}$, its signed perpendicular displacement
from waveguide $\mathcal L_n$ is given by
\begin{align}
d_n(\boldsymbol\psi)
\triangleq
\boldsymbol n^{\top}(\theta_n)\boldsymbol\psi-q_n
=
-x\sin\theta_n+y\cos\theta_n-q_n.
\label{eq:signed_line_distance}
\end{align}
Accordingly, the perpendicular distance from
$\boldsymbol\psi$ to $\mathcal L_n$ is 
$\operatorname{dist}(\boldsymbol\psi,\mathcal L_n) = |d_n(\boldsymbol\psi)|$. 

To characterize the service capability of a given deployment, let $\widetilde{\boldsymbol\psi}_n^{\rm pin}$ denote the horizontal location of the active PA on waveguide $n$. Following the commonly used spherical-wavefront LoS model for pinching-antenna systems \cite{ding2024flexible,xu2026generalized}, the equivalent channel from the feed point of waveguide $n$ to location $\boldsymbol\psi$ is given by
\begin{align}
h_n
=\frac{\sqrt{\eta}}{r_n}
\exp\!\Big[-j\Big(\frac{2\pi r_n}{\lambda}
+\frac{2\pi \ell_n(\widetilde{\boldsymbol\psi}_n^{\rm pin})}{\lambda_g}\Big)\Big],
\label{eq:pa_channel}
\end{align}
where
$r_n\triangleq\big(d_v^2+\|\boldsymbol\psi-\widetilde{\boldsymbol\psi}_n^{\rm pin}\|^2\big)^{\frac{1}{2}}$,
$\eta=c^2/(4\pi f_c)^2$, $\lambda$ and $\lambda_g$ denote the free-space and guided wavelengths, respectively, and $\ell_n(\widetilde{\boldsymbol\psi}_n^{\rm pin})$ is the guided propagation distance from the feed point to the active PA. 

Let $\boldsymbol h=[h_1,\ldots,h_N]^{\top}$ denote the channel
vector from the $N$ waveguides to the service location. The baseband
processor jointly controls the signals transmitted through the
waveguides. Under a total transmit-power constraint $P$, maximum-ratio
transmission aligns the transmitted signals with their corresponding
channel phases, yielding
\begin{align}
\Gamma(\boldsymbol\psi;
\{\widetilde{\boldsymbol\psi}_n^{\rm pin}\}_{n=1}^{N})
=
\rho\|\boldsymbol h\|^2
=
\rho\eta\sum_{n=1}^{N}\frac{1}{r_n^2},
\label{eq:instantaneous_snr}
\end{align}
where $\rho=P/\sigma^2$ and $\sigma^2$ denotes the noise power.
Since maximum-ratio transmission compensates both the free-space and
guided propagation phases in \eqref{eq:pa_channel}, the resulting SNR
depends only on the propagation distances $\{r_n\}_{n=1}^{N}$.

For a fixed waveguide deployment
$\mathcal D_N\triangleq\{\mathcal L_n\}_{n=1}^{N}$, the PA
locations are adapted to the current service location
$\boldsymbol\psi$. Since the SNR contribution of each waveguide
decreases with its propagation distance $r_n$, the optimal PA on
waveguide $n$ is the point on $\mathcal L_n$ having the minimum
Euclidean distance to $\boldsymbol\psi$. Therefore, the maximum achievable SNR at $\boldsymbol\psi$ is given by
\begin{align}
\Gamma^\star(\boldsymbol\psi;\mathcal D_N)
=
\rho\eta\sum_{n=1}^{N}
\frac{1}{d_v^2+\operatorname{dist}^2(\boldsymbol\psi,\mathcal L_n)}.
\label{eq:post_snr}
\end{align}
Since the common factor $\rho\eta$ does not affect the optimal deployment geometry, define the geometry-dependent regional value as
\begin{align}
\mathcal V_N(\mathcal D_N)
\triangleq
\min_{\boldsymbol\psi\in\mathcal A}
\sum_{n=1}^{N}
\frac{1}{d_v^2+\operatorname{dist}^2(\boldsymbol\psi,\mathcal L_n)}.
\label{eq:general_normalized_value}
\end{align}
The long-term deployment problem is therefore
\begin{align}
V_N^\star
\triangleq
\max_{\{\theta_n,q_n\}_{n=1}^{N}}
\mathcal V_N(\mathcal D_N).
\label{prob:general_deployment}
\end{align}
Problem \eqref{prob:general_deployment} jointly optimizes the orientation and offset of each waveguide. Depending on the service-region geometry and system parameters, the resulting deployment may exhibit parallel or nonparallel structures.

\begin{rem}
The infinite-line representation does not imply physically infinite
waveguides. For a waveguide with orientation $\theta$, the coordinate
of the orthogonal projection of
$\boldsymbol\psi\in\mathcal A$ onto the waveguide direction is
$\boldsymbol d^{\top}(\theta)\boldsymbol\psi$, which satisfies
\begin{align}
-U(\theta) \le \boldsymbol d^{\top}(\theta)\boldsymbol\psi
\le U(\theta),
\end{align}
where $U(\theta) = A|\cos\theta|+B|\sin\theta|$. 
Therefore, a finite waveguide segment of length $2U(\theta)$ that
covers this projection interval yields exactly the same point-to-line
distances in \eqref{eq:post_snr} for all
$\boldsymbol\psi\in\mathcal A$. In particular, a horizontal waveguide
requires a length of only $2A=D_x$. Moreover, since 
$U(\theta)\le (A^2+B^2)^{\frac{1}{2}}$, 
a length of $2(A^2+B^2)^{\frac{1}{2}}$ is sufficient for any orientation.
Hence, the infinite-line model provides an exact geometric
representation whenever the physical waveguide spans the corresponding
projection of the service region.
\end{rem}

The line-based formulation does not explicitly impose collision or
minimum-spacing constraints between different waveguides. Accordingly,
the horizontal projections of different waveguides may intersect or
overlap. Such configurations can be implemented, for example, by using
different routing layers or small vertical offsets.

The general problem \eqref{prob:general_deployment} is highly nonconvex and contains a continuum of service-location constraints. We therefore proceed in two stages. Section~\ref{sec:two_waveguide} first considers $N=2$, for which the geometry can be characterized analytically and the global role of parallel deployment can be established. Section~\ref{sec:generalN} then returns to $N\ge3$ and develops an unrestricted deployment optimization method, complemented by selected structural results.

\section{Fundamental Two-Waveguide Deployment}
\label{sec:two_waveguide}

We first specialize the general deployment problem \eqref{prob:general_deployment}
to $N=2$, which is
the fundamental setting in which the deployment geometry already becomes
nontrivial. In particular, parallel and nonparallel waveguide
configurations can lead to different regional coverage performance,
making it possible to examine whether parallel deployment is actually
optimal. In what follows, we first characterize the best parallel deployment
and then establish a sufficient condition under which it is globally
optimal among all two-waveguide configurations.

\subsection{Optimal Parallel Deployment}
\label{sec:parallel}

We first specialize \eqref{eq:general_normalized_value} to $N=2$, and write
\begin{align}
\mathcal V(\mathcal L_1,\mathcal L_2)
\!\triangleq\!
\mathcal V_2(\{\mathcal L_1,\mathcal L_2\})
\!=\!
\min_{\boldsymbol\psi\in\mathcal A}
\sum_{n=1}^{2}
\frac{1}{d_v^2\!+\!\operatorname{dist}^2(\boldsymbol\psi,\mathcal L_n)}.
\label{eq:normalized_value}  
\end{align}
Let $V^\star\triangleq V_2^\star$ denote the optimal value of the
general two-waveguide deployment problem, and let $V_{\rm P}^\star$
denote the optimal value when the two waveguides are restricted to be
parallel.

Consider two parallel lines with common direction
$\boldsymbol d(\theta)$ and normal $\boldsymbol n(\theta)$, 
as shown in Fig. \ref{fig:parallel_projection}(b.1).
Let $q_1$ and $q_2$ denote their signed coordinates along
the common normal direction. For a service location
$\boldsymbol\psi=[x,y]^{\mathsf T}\in\mathcal A$, define
\begin{align}
t
\triangleq
\boldsymbol n^{\mathsf T}(\theta)\boldsymbol\psi
=
-x\sin\theta+y\cos\theta .
\label{eq:normal_coordinate}
\end{align}
The distance from $\boldsymbol\psi$ to waveguide
$\mathcal L_n$ is therefore $|t-q_n|$. Moreover, as
$\boldsymbol\psi$ varies over the rectangular service region,
its normal coordinate spans
\begin{align}
-W(\theta)\le t\le W(\theta),
\label{eq:projection_width}
\end{align}
where $W(\theta) = A|\sin\theta|+B|\cos\theta|$.

As illustrated in Fig.~\ref{fig:parallel_projection}(b.2), for a fixed
common orientation $\theta$, the original two-dimensional
parallel-waveguide deployment problem is reduced exactly to a
one-dimensional placement problem for $q_1$ and $q_2$ over the
projected interval $[-W(\theta),W(\theta)]$.

The common orientation therefore affects the parallel deployment
problem only through the width of this projected interval. A larger
$W(\theta)$ enlarges the set of service locations over which the
worst-case performance is evaluated. For any given $q_1$ and $q_2$,
the minimum received-power sum over a larger interval cannot exceed
that over a smaller interval. This remains true after optimizing
$q_1$ and $q_2$. Hence, the optimal parallel orientation is obtained
by minimizing $W(\theta)$. Since $A\ge B$, we have
\begin{align}
    W(\theta) = A|\sin\theta|+B|\cos\theta| \ge B,
\end{align}
with equality at $\theta=0$. Therefore, an optimal parallel
deployment can be chosen horizontal, i.e., parallel to the longer
side of the service region. 

It remains to optimize the vertical positions of the two horizontal
waveguides. Since $\theta=0$, the normal coordinate in
\eqref{eq:normal_coordinate} reduces to the vertical coordinate $y$.
The following lemma characterizes the optimal midpoint for a fixed waveguide separation.

\begin{lem}
\label{lem:parallel_centering}
An optimal pair of horizontal waveguides can be chosen with both
waveguide positions in $[-B,B]$. Writing their positions as
$m-s$ and $m+s$, for every fixed $s\in[0,B]$, the optimal midpoint
is $m=0$. Hence, the two waveguides can be written as
\begin{align}
    \mathcal L_1:y=-s,\ \mathcal L_2:y=s,\ 0\le s\le B.
\end{align}
Moreover, for a given $s$, the worst service location is either
$y=0$ or $y=\pm B$.
\end{lem}

   {\bf Proof:} See Appendix \ref{appd: lemma vertical coordinate}.
   \hfill $\blacksquare$

By Lemma~\ref{lem:parallel_centering}, it remains only to optimize
the half-separation $s$. Define
\begin{align}
    G_{\rm c}(s) &\triangleq g_s(0) = \frac{2}{d_v^2+s^2},  \label{eq:center_value}\\
    G_{\rm e}(s) &\triangleq g_s(B) =\frac{1}{d_v^2+(B-s)^2} +\frac{1}{d_v^2+(B+s)^2}.
    \label{eq:edge_value}
\end{align}
The worst-location value can therefore be written as
    \begin{align}
        V_{\rm P}(s) = \min\{G_{\rm c}(s),G_{\rm e}(s)\}.
    \label{eq:parallel_objective}
\end{align}

We next characterize how the two candidate values vary with $s$.
The center value $G_{\rm c}(s)$ decreases monotonically as the
waveguides move away from the centerline. Moreover,
\begin{align}
G_{\rm c}(s)-G_{\rm e}(s)
=
\frac{
2B^2(B^2+d_v^2-3s^2)
}{
(d_v^2+s^2)
[d_v^2+(B-s)^2]
[d_v^2+(B+s)^2]
}.
\label{eq:Gc_Ge_difference}
\end{align}
Hence, the edge is the limiting location when
$s^2<(B^2+d_v^2)/3$, whereas the center becomes limiting when
$s^2>(B^2+d_v^2)/3$. When the two values can be balanced, their
intersection occurs at
\begin{align}
s_{\rm eq}^2
=
\frac{B^2+d_v^2}{3}.
\label{eq:s_eq}
\end{align}

Since $G_{\rm e}(s)$ is the limiting value for $s<s_{\rm eq}$,
whether the optimum occurs at the balancing point $s_{\rm eq}$
depends on how $G_{\rm e}(s)$ varies with $s$. Differentiating $G_{\rm e}(s)$ gives
\begin{align}
\frac{{\rm d}G_{\rm e}(s)}{{\rm d}s}
=
-\frac{
4s\left[
(s^2+B^2+d_v^2)^2
-4B^2(B^2+d_v^2)
\right]
}{
[d_v^2+(B-s)^2]^2
[d_v^2+(B+s)^2]^2
}.
\label{eq:Ge_derivative}
\end{align}
For $s>0$, the sign of the derivative is therefore determined by
\begin{align}
H(s)
\triangleq
(s^2+B^2+d_v^2)^2
-4B^2(B^2+d_v^2),
\end{align}
which increases monotonically with $s$. At $s=0$,
\begin{align}
H(0)
=
(B^2+d_v^2)(d_v^2-3B^2).
\end{align}
Hence, when $d_v^2<3B^2$, we have $H(0)<0$ and $H(s)$
crosses zero once as $s$ increases. Setting
$H(s)=0$ gives the unique positive stationary point
\begin{align}
s_{\rm e}^2
=
2B\sqrt{B^2+d_v^2}-(B^2+d_v^2).
\label{eq:s_edge}
\end{align}
Since $H(s)<0$ for $0<s<s_{\rm e}$ and $H(s)>0$ for
$s>s_{\rm e}$, \eqref{eq:Ge_derivative} shows that
$G_{\rm e}(s)$ increases up to $s_{\rm e}$ and decreases
thereafter. Thus, $s_{\rm e}$ is the maximizer of the edge value.
When $d_v^2\ge3B^2$, we have $H(s)\ge0$ for all $s\ge0$,
so $G_{\rm e}(s)$ decreases with $s$ and is maximized at $s=0$.

For $d_v^2<3B^2$, we now compare the maximizer $s_{\rm e}$ of
the edge value with the balancing point $s_{\rm eq}$ at which
$G_{\rm e}(s)=G_{\rm c}(s)$. Direct comparison gives
\begin{align}
s_{\rm e}=s_{\rm eq}
\ \Longleftrightarrow\ 
d_v^2=\frac{5}{4}B^2.
\end{align}
When $d_v^2<5B^2/4$, we have $s_{\rm eq}<s_{\rm e}$.
In this case, $G_{\rm e}(s)$ is still increasing when it meets
the decreasing $G_{\rm c}(s)$. The worst-location value is
therefore maximized at their intersection, giving
$s_{\rm P}^\star=s_{\rm eq}$. When
$5B^2/4<d_v^2<3B^2$, we have $s_{\rm e}<s_{\rm eq}$.
Thus, $G_{\rm e}(s)$ reaches its maximum before meeting
$G_{\rm c}(s)$ and, since it is still the limiting value,
$s_{\rm P}^\star=s_{\rm e}$.

Finally, when $d_v^2\ge3B^2$, we have
$s_{\rm eq}>B$, so the edge value is the limiting value throughout
the feasible interval $0\le s\le B$. Since $G_{\rm e}(s)$
decreases with $s$ in this regime, the optimum is attained at
$s_{\rm P}^\star=0$.

Combining these cases yields
\begin{align}
s_{\rm P}^\star
=
\begin{cases}
\displaystyle
\sqrt{\frac{B^2+d_v^2}{3}},
& d_v^2\le\frac54B^2,\\[1.5mm]
\displaystyle
\sqrt{2B\sqrt{B^2+d_v^2}-(B^2+d_v^2)},
& \frac54B^2<d_v^2<3B^2,\\[1.5mm]
0,
& d_v^2\ge3B^2,
\end{cases}
\label{eq:sp_star}
\end{align}
with the corresponding optimal value
\begin{align}
V_{\rm P}^\star
=
\begin{cases}
\displaystyle
\frac{6}{B^2+4d_v^2},
& d_v^2\le\frac54B^2,\\[1.5mm]
\displaystyle
\frac{B+\sqrt{B^2+d_v^2}}{2Bd_v^2},
& \frac54B^2<d_v^2<3B^2,\\[1.5mm]
\displaystyle
\frac{2}{B^2+d_v^2},
& d_v^2\ge3B^2.
\end{cases}
\label{eq:vp_star}
\end{align}

\subsection{Global Optimality of Parallel Deployment}
\label{sec:global_optimality}

The preceding subsection characterizes the optimal deployment within
the parallel class. We now turn to the general two-waveguide problem and establish a
sufficient condition under which the parallel solution remains
globally optimal.

\begin{prop}\label{prop:global_parallel}
Let $\kappa\triangleq A/B\geq1$ and define
\begin{align}
\tau(\kappa)
\triangleq
\min\left\{
\frac{5}{4},
\frac{2\kappa^2-1}{\kappa^2+1}
\right\},
\ \kappa\geq1.
\label{eq:tau_kappa}
\end{align}
If $\frac{d_v^2}{B^2}\leq\tau(\kappa)$, 
then a globally optimal solution to the unrestricted
two-waveguide deployment problem can be chosen as
\begin{align}
    \mathcal L_1^\star  : y=-s^\star,\ 
    \mathcal L_2^\star : y=s^\star,\ 
    s^\star = \sqrt{\frac{B^2+d_v^2}{3}},
\end{align}
with corresponding optimal worst-location value $V^\star = \frac{6}{B^2+4d_v^2}$. 
\end{prop}

{\bf Proof:}
    The key idea behind Proposition~\ref{prop:global_parallel} is to
    normalize the service region by $B$ and exploit the largest
    axis-aligned ellipse contained in the rectangle. For an arbitrary
    pair of lines, we identify suitably chosen boundary points of this
    ellipse and derive an upper bound on the corresponding worst-location
    received-power sum. The resulting bound depends on the normalized
    waveguide height $d_v/B$ and the aspect ratio $A/B$, leading to the
    condition $\frac{d_v^2}{B^2}\leq\tau(\kappa)$. The detailed proof is
    provided in Appendix~\ref{app:prop1}.
    \hfill$\blacksquare$

The two-waveguide analysis provides a useful benchmark for the general
problem. It shows that parallel deployment is globally optimal in a
certain parameter regime, while nonparallel configurations can provide
better regional coverage outside this regime. These results also 
suggest that the waveguide geometry may benefit from
being optimized rather than fixed a priori. We next extend the
deployment design to systems with more than two waveguides.

\section{General Multi-Waveguide Deployment Optimization}
\label{sec:generalN}

The two-waveguide analysis establishes an explicit geometry-dependent
regime in which a parallel architecture is globally optimal. We now
turn to the general case with $N\ge3$ waveguides. Compared with the
two-waveguide case, obtaining comparable analytical guarantees becomes
considerably more difficult because the orientations and offsets of
multiple waveguides jointly determine the regional coverage. We
therefore directly optimize the general deployment in
\eqref{prob:general_deployment} without imposing a predefined
orientation pattern. We first formulate the resulting multi-waveguide
optimization problem and develop an exchange-based algorithm for its
solution. We then consider the practically relevant parallel-waveguide
case and derive a simple structural guideline for its deployment.

\subsection{General Multi-Waveguide Formulation}
\label{subsec:generalN_structure}

Collect the deployment variables as
\begin{align}
\boldsymbol\Theta
 \triangleq[\theta_1,\ldots,\theta_N]^{\mathsf T}, \
\boldsymbol q
 \triangleq[q_1,\ldots,q_N]^{\mathsf T},
\end{align}
and define the geometry-dependent normalized received-power sum at service
location $\boldsymbol\psi$ as
\begin{align}
    F_N(\boldsymbol\psi;\boldsymbol\Theta,\boldsymbol q) \triangleq
    \sum_{n=1}^{N} \frac{1}{d_v^2+d_n^2(\boldsymbol\psi)}.
    \label{eq:N_received_sum}
\end{align}
The deployment problem in \eqref{prob:general_deployment} can then be
written as
\begin{align}
    V_N^\star =
    \max_{\boldsymbol\Theta,\boldsymbol q}
    \; \min_{\boldsymbol\psi\in\mathcal A}
    F_N(\boldsymbol\psi;\boldsymbol\Theta,\boldsymbol q).
    \label{eq:N_original_problem}
\end{align}
As seen, no common orientation is imposed in \eqref{eq:N_original_problem}, so
parallel, crossing, and mixed-orientation deployments are all included
in the feasible set.

Before developing the optimization method, we first identify a compact
search region for the waveguide offsets. For a given orientation
$\theta$, the projection of $\mathcal A$ onto the corresponding normal
direction lies in  $[-W(\theta),W(\theta)]$ with $W(\theta) \triangleq A|\sin\theta|+B|\cos\theta|$. 
If $q_n>W(\theta_n)$, moving waveguide $n$ to
$q_n=W(\theta_n)$ weakly decreases its distance to every service
location in $\mathcal A$. The same argument applies when
$q_n<-W(\theta_n)$. Hence, without loss of optimality, we have
\begin{align}
    |q_n| \le W(\theta_n) \le \sqrt{A^2+B^2}, \ n=1,\ldots,N.
    \label{eq:N_compact_domain}
\end{align}
Meanwhile, defining $R\triangleq\sqrt{A^2+B^2}$, 
the search can therefore be restricted to
\begin{align}
    0\le\theta_n\le\pi, \ -R\le q_n\le R, \ n=1,\ldots,N.
\label{eq:N_box_domain}
\end{align}

Problem \eqref{eq:N_original_problem} remains highly nonconvex, and
its objective is determined by the worst service location over the
continuous region $\mathcal A$. We therefore develop an adaptive
optimization method that jointly updates all waveguide orientations
and offsets while progressively identifying the service locations that
are most critical to the regional max--min objective.

\subsection{Waveguide Deployment Optimization}
\label{subsec:N_algorithm}

Problem \eqref{eq:N_original_problem} is a semi-infinite max--min
problem with only $2N$ deployment variables but a continuum of
service locations determining the worst-case performance. A
straightforward approach is to discretize the service region into a
finite set of candidate locations and solve the resulting
finite-dimensional problem, as adopted in
\cite{xu2026environment1}. However, a dense
discretization can make each deployment update computationally
expensive, whereas a coarse grid may fail to capture the true
worst-served location.

Exchange methods provide a classical approach to semi-infinite
programming by solving a finite restriction of the original problem
and progressively adding newly identified critical points
\cite{blankenship1976infinitely,hettich1993semi}. Following this
principle, we develop an exchange-based algorithm tailored to the
present waveguide deployment problem. The algorithm alternates
between optimizing the deployment over a small working set of service
locations and searching the continuous region for a newly identified
worst-served point.

Let $\mathcal S^{(r)} = \{\boldsymbol\psi_1,\ldots,\boldsymbol\psi_{M_r}\}
\subset\mathcal A $
denote the working set at exchange iteration $r$, and define 
$F_m(\boldsymbol\Theta,\boldsymbol q)
\triangleq
F_N(\boldsymbol\psi_m;\boldsymbol\Theta,\boldsymbol q)$. 
For a fixed working set, the corresponding objective
$\min_m F_m$ is nonsmooth. We therefore optimize its smooth lower
approximation
\begin{align}
\widetilde V_{\beta}
(\boldsymbol\Theta,\boldsymbol q;\mathcal S^{(r)})
\triangleq
-\frac{1}{\beta}
\log\!\left[
\sum_{m=1}^{M_r}
\exp\!\left(-\beta F_m(\boldsymbol\Theta,\boldsymbol q)\right)
\right],
\label{eq:N_softmin}
\end{align}
which satisfies
\begin{align}
\min_mF_m-\frac{\log M_r}{\beta}
\le
\widetilde V_{\beta}
\le
\min_mF_m.
\label{eq:N_softmin_bound}
\end{align}
Thus, gradually increasing $\beta$ provides an increasingly accurate
approximation of the max--min objective over the current working set.

The smooth objective in \eqref{eq:N_softmin} can be efficiently
optimized because its gradient is available in closed form. For
$\boldsymbol\psi_m=[x_m,y_m]^{\mathsf T}$, using the signed
displacement $d_n(\boldsymbol\psi_m)$ defined in
\eqref{eq:signed_line_distance}, we have
\begin{align}
\frac{\partial F_m}{\partial q_n}
&=
\frac{2d_n(\boldsymbol\psi_m)}
{\left[d_v^2+d_n^2(\boldsymbol\psi_m)\right]^2},
\label{eq:N_gradient_q}\\
\frac{\partial F_m}{\partial\theta_n}
&=
\frac{
2d_n(\boldsymbol\psi_m)
(x_m\cos\theta_n+y_m\sin\theta_n)
}{
\left[d_v^2+d_n^2(\boldsymbol\psi_m)\right]^2
}.
\label{eq:N_gradient_theta}
\end{align}
Define the soft-min weights as
\begin{align}
\omega_m
\triangleq
\frac{
\exp(-\beta F_m)
}{
\sum_{j=1}^{M_r}\exp(-\beta F_j)
}.
\label{eq:N_softmin_weights}
\end{align}
The gradient of \eqref{eq:N_softmin} is then
\begin{align}
\nabla\widetilde V_{\beta}
=
\sum_{m=1}^{M_r}
\omega_m\nabla F_m.
\label{eq:N_softmin_gradient}
\end{align}

The weights in \eqref{eq:N_softmin_weights} also provide an
interpretation of the smoothing. Let
\begin{align}
\mathcal I(\boldsymbol\Theta,\boldsymbol q)
\triangleq
\big\{
m:
F_m(\boldsymbol\Theta,\boldsymbol q)
=
\min_j F_j(\boldsymbol\Theta,\boldsymbol q)
\big\}
\label{eq:N_active_worst_set}
\end{align}
denote the set of worst-served locations within the current working
set. As $\beta$ increases, the total weight assigned to indices
outside $\mathcal I$ vanishes, and the smooth gradient approaches a
convex combination of the gradients associated with these
worst-served locations. Hence, at an interior limiting stationary
point, the continuation procedure approaches
\begin{align}
\sum_{m\in\mathcal I}
\lambda_m\nabla F_m
=
\boldsymbol 0, \ \
\lambda_m\ge0, \ \
\sum_{m\in\mathcal I}\lambda_m=1,
\label{eq:N_stationarity_balance}
\end{align}
with the corresponding projected first-order condition when a
deployment variable lies on the boundary. This condition indicates
that the optimized deployment balances the competing spatial
gradients associated with multiple critical service locations.

For each fixed working set and smoothing parameter, we optimize
\eqref{eq:N_softmin} over the compact domain in
\eqref{eq:N_box_domain} using projected gradient ascent with Armijo
backtracking \cite{bertsekas1976goldstein}. The inner iterations are terminated when the
projected-gradient norm falls below a prescribed tolerance. We then
increase $\beta$ and warm-start the next continuation stage from the
previous solution. 

After each deployment update, we search the continuous service region
for the current worst-served location according to
\begin{align}
\boldsymbol\psi_{\rm w}^{(r)}
\in
\arg\min_{\boldsymbol\psi\in\mathcal A}
F_N\!\left(
\boldsymbol\psi;
\boldsymbol\Theta^{(r)},\boldsymbol q^{(r)}
\right).
\label{eq:N_worst_location}
\end{align}
The identified point is then added to the working set as
\begin{align}
\mathcal S^{(r+1)}
=
\mathcal S^{(r)}
\cup
\{\boldsymbol\psi_{\rm w}^{(r)}\},
\end{align}
and the exchange iterations continue until the difference between
the worst-location value over $\mathcal A$ and that over the current
working set falls below a prescribed tolerance.
Since this search involves only the two-dimensional service location,
it can be efficiently performed by evaluating a moderate rectangular
grid and locally refining the lowest-valued grid points.

The grid resolution can also be related to the accuracy of the
worst-location search. For $f(u)=\frac{1}{d_v^2+u^2}$, 
we have
\begin{align}
    |f'(u)|
    =
    \frac{2|u|}{(d_v^2+u^2)^2}
    \leq
    \frac{3\sqrt{3}}{8d_v^3},
\end{align}
where the maximum is attained at
$|u|=d_v/\sqrt{3}$. Moreover, since
$d_n(\boldsymbol\psi)
=\mathbf n^T(\theta_n)\boldsymbol\psi-q_n$ and
$\|\mathbf n(\theta_n)\|=1$, the chain rule gives 
    $\left\|
    \nabla_{\boldsymbol\psi}
    f\big(d_n(\boldsymbol\psi)\big)
    \right\|
    \leq
    \frac{3\sqrt{3}}{8d_v^3}$. 
Therefore,
\begin{align}
    \left\|
    \nabla_{\boldsymbol\psi}
    F_N(\boldsymbol\psi;\boldsymbol\Theta,\mathbf q)
    \right\|
    \leq
    \frac{3\sqrt{3}N}{8d_v^3}.
\end{align}
Hence, $F_N(\boldsymbol\psi;\boldsymbol\Theta,\mathbf q)$ is
Lipschitz continuous over $\mathcal A$ with constant  
$L_N = \frac{3\sqrt{3}N}{8d_v^3}$. 
If $\mathcal G$ is a grid with covering radius
$\delta_{\mathcal G}$, then
\begin{align}
\min_{\boldsymbol g\in\mathcal G}
F_N(\boldsymbol g)
-
L_N\delta_{\mathcal G}
\le
\min_{\boldsymbol\psi\in\mathcal A}
F_N(\boldsymbol\psi)
\le
\min_{\boldsymbol g\in\mathcal G}
F_N(\boldsymbol g).
\label{eq:N_grid_certificate}
\end{align}
Hence, the grid search provides a certified lower bound on the
continuous-region worst-location value, with an accuracy controlled
by $\delta_{\mathcal G}$.

\subsection{Optimal Parallel Multi-Waveguide Deployment}
\label{subsec:N_parallel_guidelines}

Although the general formulation allows different waveguide
orientations, parallel deployment may be preferred in practice because
of its simple installation and regular geometry. In this case, the
deployment problem can be considerably simplified.
Let all $N$ waveguides have a common orientation $\theta$, with signed
offsets $\{q_n\}_{n=1}^{N}$ along the common normal direction. The
optimal parallel deployment is defined as
\begin{align}
V_{{\rm P},N}^\star
\triangleq
\max_{\theta,\{q_n\}_{n=1}^{N}}
\;
\min_{\boldsymbol\psi\in\mathcal A}
\sum_{n=1}^{N}
\frac{1}{
d_v^2+
[\boldsymbol n^{\mathsf T}(\theta)\boldsymbol\psi-q_n]^2
}.
\label{eq:N_parallel_general}
\end{align}

\begin{lem}
\label{lem:N_parallel_guideline}
Consider $N\ge2$ parallel waveguides over
$\mathcal A=[-A,A]\times[-B,B]$ with $A\ge B$.
An optimal common orientation can be chosen parallel to the longer
side of the service region. Hence, the parallel deployment problem
reduces to
\begin{align}
V_{{\rm P},N}^\star
=
\max_{\substack{\{q_n\}_{n=1}^{N}:\\ |q_n|\le B}}
\;
\min_{|y|\le B}
\sum_{n=1}^{N}
\frac{1}{d_v^2+(y-q_n)^2}.
\label{eq:N_parallel_1D_guideline}
\end{align}
\end{lem}

{\bf Proof:}
The proof follows the same projection argument used for the two-waveguide parallel deployment. Specifically, for a common orientation $\theta$, define
$t=\boldsymbol n^{\mathsf T}(\theta)\boldsymbol\psi$. The rectangular
service region can be projected onto 
${\small{-W(\theta) \le t\le W(\theta)}}$ with {\small{$W(\theta) = A|\sin\theta|+B|\cos\theta|$}}.  
For any fixed waveguide offsets, enlarging this interval cannot
increase the worst-location value. Therefore, an optimal common
orientation is obtained by minimizing $W(\theta)$. Since $A\ge B$,
we have $W(\theta)\ge B$, with equality at $\theta=0$. Thus,
an optimal parallel deployment can be chosen along the longer side
of the service region. With $\theta=0$, the normal coordinate reduces
to $y$, which directly gives \eqref{eq:N_parallel_1D_guideline}.
\hfill $\blacksquare$

Lemma~\ref{lem:N_parallel_guideline} provides a simple procedure for
practical environments in which parallel waveguides are required.
The waveguides are first aligned with the longer side of the service
region, so no orientation search is needed. The remaining task is to
determine only their transverse offsets $\{q_n\}$ over $[-B,B]$.

A simple coordinate-search procedure can be used for this purpose.
Starting from an initial set of offsets, each $q_n$ is optimized in
turn over $[-B,B]$ while keeping the other offsets fixed. After all
waveguides are updated, the procedure is repeated until the
worst-location value in \eqref{eq:N_parallel_1D_guideline} no longer
improves noticeably. Since each update involves only a one-dimensional
search, the resulting deployment procedure is simple to implement.

\subsection{Large-Height Deployment Structure}

The optimized multi-waveguide geometry may exhibit different orientation
patterns at finite waveguide heights. Nevertheless, a simple structural
property can be obtained in the large-height regime. Importantly, this
result concerns the general deployment problem and does not impose a
parallel architecture.

\begin{lem}
\label{lem:large_height}
Consider a nonsquare service region with $A>B$, and let
$\mathcal{D}_N^\star(d_v)$ denote a globally optimal solution of the
general $N$-waveguide deployment problem. As $d_v/B\rightarrow\infty$,
every accumulation point of $\mathcal{D}_N^\star(d_v)$ consists of
coincident centerline waveguides parallel to the longer side of the
service region. Equivalently,
\begin{align}
    |q_n^\star| \rightarrow 0, \ |\sin\theta_n^\star| \rightarrow 0,
    \ \forall n=1,\ldots,N .
\end{align}
\end{lem}

{\bf Proof:}
    See appendix \ref{appd: large height}. 
\hfill $\blacksquare$
 
Lemma~\ref{lem:large_height} shows that parallelism can emerge from the
general optimization rather than being imposed a priori. At finite
heights, different waveguide orientations may be beneficial, whereas
for sufficiently large $d_v/B$ the optimized geometry approaches
coincident centerline waveguides aligned with the longer side. This property will be validated by the simulation results in the following numerical results section.

\begin{rem} 
    Sufficiently long waveguides and unconstrained deployment areas are assumed in this work to focus on a basic setting and simplify the geometric characterization. In practical installations, however, finite waveguide lengths and restricted deployment regions may limit the feasible waveguide locations, orientations, and PA positions. Incorporating such geometric constraints into the deployment design is therefore an important extension of the present framework.
\end{rem}

\begin{rem}
    The worst-location service performance is adopted as the deployment objective to ensure robust coverage over the entire service region. Other long-term objectives are also of interest. For example, when the spatial user distribution is known, the waveguide geometry can instead be optimized to maximize the expected achievable rate over the region. Such objectives may lead to different preferred deployment structures and deserve further investigation.
\end{rem}

\section{Numerical Results}
\label{sec:numerical}

We present numerical results to verify the analytical findings, validate the
proposed adaptive exchange algorithm, and investigate the waveguide
deployment behavior beyond the theoretically characterized regimes. We
consider a carrier frequency of $f_c=28$ GHz and an effective refractive
index of $n_{\rm eff}=1.4$. We use $B=10$ m as the
reference spatial scale and set $A=\kappa B$, where
$\kappa\triangleq A/B\in\{1,1.5,2,3\}$. Each waveguide is assumed to be
sufficiently long to cover the projection of the service region.
Since the common factor $\rho\eta$ does not affect the
optimal deployment geometry, we evaluate the geometry-dependent regional
value defined in \eqref{eq:general_normalized_value}. The results are
therefore presented primarily in terms of the normalized waveguide height
$d_v/B$.

\begin{figure}[t]
    \centering
    \includegraphics[width=0.48\columnwidth]{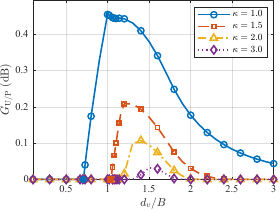}          \captionsetup{justification=justified, singlelinecheck=false, font=small}	      
    \caption{Performance gain of the unrestricted two-waveguide
    deployment over the optimal parallel deployment versus the normalized
    waveguide height $d_v/B$ for different region aspect ratios
    $\kappa=A/B$. The filled markers on the horizontal axis indicate the
    sufficient global-optimality thresholds
    $d_v/B=\sqrt{\tau(\kappa)}$ in Proposition~\ref{prop:global_parallel}.} \vspace{-5mm}
    \label{fig:gain_unrestricted_parallel}
\end{figure}

\subsection{Two-Waveguide Performance}

We first examine the performance gain obtained by allowing arbitrary
waveguide orientations relative to the optimal parallel deployment.
Define
\begin{align}
G_{\rm U/P}
\triangleq
10\log_{10}
\left(
\frac{V_{2}^{\rm num}}
{V_{\rm P}^{\star}}
\right),
\label{eq:gain_unrestricted_parallel}
\end{align}
where $V_{2}^{\rm num}$ denotes the numerically obtained objective
value of the unrestricted two-waveguide problem, and
$V_{\rm P}^{\star}$ is given in \eqref{eq:vp_star}.

As shown in Fig.~\ref{fig:gain_unrestricted_parallel}, the
unrestricted deployment provides no gain over the optimal parallel
deployment throughout the global-optimality regime guaranteed by
Proposition~\ref{prop:global_parallel}. The filled markers on the
horizontal axis indicate the corresponding sufficient thresholds
$d_v/B=\sqrt{\tau(\kappa)}$. For
$\kappa=1$, $1.5$, $2$, and $3$, these thresholds are approximately
$0.707$, $1.038$, $1.118$, and $1.118$, respectively. Beyond the
guaranteed regimes, the behavior depends strongly on the service-region
aspect ratio.

For the square case $\kappa=1$, a noticeable gain appears immediately
after the sufficient threshold and rapidly increases to a peak of about
$0.45$ dB. As $\kappa$ increases, the gain becomes both weaker and more
delayed. In particular, for $\kappa=1.5$, the peak gain is about
$0.2$ dB, while for $\kappa=2$ it further decreases to around
$0.1$ dB. For $\kappa=3$, only a very small improvement is observed
over a narrow range of normalized waveguide heights. These results
show that incorporating the aspect ratio into
Proposition~\ref{prop:global_parallel} substantially enlarges the
theoretically guaranteed parallel-optimality regime compared with an
aspect-ratio-independent characterization. They also suggest that the
sufficient condition can remain conservative for elongated service
regions, where the optimized unrestricted and parallel deployments
exhibit nearly identical performance beyond the guaranteed regime.
Overall, the preferred two-waveguide geometry can transition from
parallel to nonparallel as $d_v/B$ increases, with this transition
becoming less pronounced and occurring at larger heights for more
elongated service regions.



\vspace{-3mm}

\subsection{Validation of the Multi-Waveguide Algorithm}

Before investigating the general multi-waveguide deployment behavior, we first assess the solution quality of the proposed adaptive exchange algorithm for the representative case of $N=3$. Since the underlying deployment problem is nonconvex, the proposed algorithm does not guarantee global optimality. We therefore construct a branch-and-bound (BnB) benchmark over a finite service-location grid to obtain a certified upper bound on the globally optimal objective value.
Specifically, let $\mathcal{G}\subset\mathcal{A}$ denote the grid used by
the BnB procedure, and define
\begin{equation}
    V_{\mathcal{G}}^\star
    \triangleq
    \max_{\bm{\Theta},\bm{q}}
    \min_{\bm{\psi}\in\mathcal{G}}
    F_3(\bm{\psi};\bm{\Theta},\bm{q}).
\end{equation}
Because $\mathcal{G}\subset\mathcal{A}$, restricting the worst-location
search to $\mathcal{G}$ yields $ V_3^\star \leq V_{\mathcal{G}}^\star$. 
Let $V_{\rm Ex}$ denote the continuous-region objective value achieved by
the proposed algorithm, and let $\overline{V}_{\rm BnB}$ denote the global upper bound
returned by BnB. We then have $V_{\rm Ex} \leq V_3^\star \leq V_{\mathcal{G}}^\star \leq \overline{V}_{\rm BnB}$. 
Accordingly, we evaluate the certified optimality-gap bound as
\begin{equation}
    G_{\rm cert}
    \triangleq
    10\log_{10}
    \left(
    \frac{\overline{V}_{\rm BnB}}
         {V_{\rm Ex}}
    \right).
\end{equation}

Table~\ref{tab:bnb_runtime} further compares the proposed algorithm with
the BnB benchmark in terms of both solution quality and computational cost.
The BnB procedure is initialized using the solution obtained by the proposed
algorithm, while the reported BnB runtime excludes the time required to
generate this initial solution. Hence, the runtime comparison is conservative
in favor of BnB. As shown in the table, the certified optimality-gap bound
remains below $0.08$~dB for all considered aspect ratios, while the proposed
algorithm requires only about $0.15$--$0.18$~s, compared with
$27.5$--$33.3$~s for BnB. This demonstrates that the proposed algorithm
achieves a favorable solution-quality--complexity tradeoff.

\begin{table}[t]
    \centering \tiny
    \captionsetup{justification=justified, singlelinecheck=false, font=small}
    \caption{Comparison with the BnB benchmark for $N=3$ and $d_v/B=1.5$.}
    \label{tab:bnb_runtime}
    \renewcommand{\arraystretch}{1}
    \resizebox{\columnwidth}{!}{
    \begin{tabular}{c|cc|c|cc} 
        \hline \hline
        \multirow{2}{*}{$\kappa$}
        & \multicolumn{2}{c|}{Normalized max--min SNR}
        & \multirow{2}{*}{Gap [dB]}
        & \multicolumn{2}{c}{Runtime [s]} \\
        & Proposed & BnB
        & & Proposed & BnB \\
        \hline
        1.0 & 0.987 & 1.005 & 0.077 & 0.18 & 33.29 \\
        1.5 & 0.956 & 0.969 & 0.062 & 0.15 & 30.84 \\
        2.0 & 0.944 & 0.959 & 0.068 & 0.17 & 29.59 \\
        3.0 & 0.935 & 0.952 & 0.078 & 0.17 & 27.53 \\
        \hline
        \hline
    \end{tabular}}
\end{table}

\vspace{-3mm}

\subsection{Multi-Waveguide Performance}

We next investigate the performance benefit of optimizing the general
multi-waveguide geometry relative to the best parallel deployment. For
$N\geq 3$, define
\begin{equation}
    G_{{\rm U/P},N}
    \triangleq
    10\log_{10}
    \left(
    \frac{V_{{\rm U},N}}
         {V_{{\rm P},N}^{\star}}
    \right),
\end{equation}
where $V_{{\rm U},N}$ denotes the objective value achieved by
the proposed algorithm without imposing a common waveguide orientation, and
$V_{{\rm P},N}^{\star}$ denotes the optimal value under the parallel
deployment constraint.

\begin{figure}[t]
    \centering
    \includegraphics[width=0.48\columnwidth]{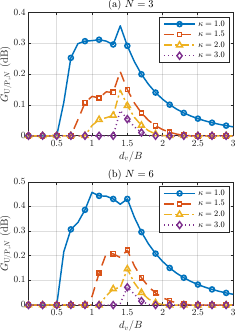}
    \captionsetup{justification=justified, singlelinecheck=false, font=small}
    \caption{Performance gain of the optimized general deployment over
    the optimal parallel deployment versus the normalized waveguide
    height $d_v/B$ for different region aspect ratios $\kappa$.} \vspace{-3mm}
    \label{fig:multi_waveguide_parallel}
\end{figure}

As shown in Fig.~\ref{fig:multi_waveguide_parallel}, the
multi-waveguide results exhibit a trend similar to that observed for the
two-waveguide case. In particular, the benefit of allowing different
waveguide orientations is most pronounced for near-square service
regions and becomes progressively smaller as the aspect ratio $\kappa$
increases. The gain is also mainly observed over an intermediate range
of normalized waveguide heights, whereas the optimized parallel
deployment remains highly competitive at relatively small and large
$d_v/B$.
Comparing the two panels, the benefit of general deployment is more
pronounced for $N=6$ than for $N=3$ in the considered settings,
especially for smaller aspect ratios. This indicates that, when more
waveguides are available, jointly optimizing their individual
orientations can provide additional geometric flexibility. Nevertheless,
for elongated service regions, the gain over the optimal parallel
deployment remains relatively limited, supporting the parallel
architecture as an attractive low-complexity deployment option in such
scenarios.

To further evaluate the benefit of optimizing the transverse
waveguide locations within the parallel architecture, we compare the
optimized parallel deployment with a commonly adopted equally spaced
parallel deployment \cite{qian2026pinching,xu2026environment1}.
Denote the corresponding worst-location value by
$V_{{\rm Eq},N}$. We define the gain of the optimized parallel
deployment over equal spacing as
\begin{align}
    G_{{\rm P/Eq},N}
    \triangleq
    10\log_{10}
    \left(
    \frac{V_{{\rm P},N}^{\star}}
         {V_{{\rm Eq},N}}
    \right).
    \label{eq:parallel_equal_gain}
\end{align}
Since both deployments are parallel and aligned with the longer side
of the service region, their worst-location values depend only on
$B$, $d_v$, and $N$, and are therefore independent of the aspect
ratio $\kappa$.

\begin{figure}[t]
    \centering
    \includegraphics[width=0.48\linewidth]{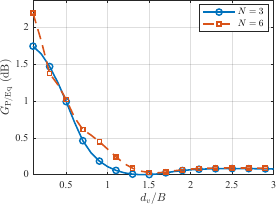}
    \captionsetup{justification=justified, singlelinecheck=false, font=small}
    \caption{Performance gain of the optimized parallel deployment over
    the equally spaced parallel deployment versus the normalized
    waveguide height $d_v/B$ for $N=3$ and $N=6$.} \vspace{-3mm}
    \label{fig:parallel_equal_gain}
\end{figure}

\begin{figure*}[t]
    \centering
    \includegraphics[width=0.92\textwidth]{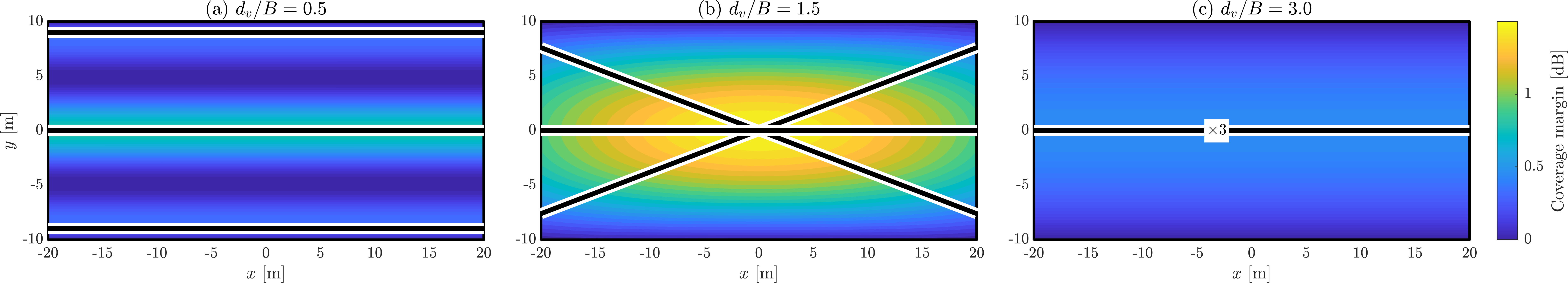}
    \captionsetup{justification=justified, singlelinecheck=false, font=small}
    \caption{Representative optimized three-waveguide deployments for a service region with $B=10$ m and $\kappa=A/B=2$.} \vspace{-3mm}
    \label{fig:N3_optimal_geometries}
\end{figure*}

As shown in Fig.~\ref{fig:parallel_equal_gain}, optimizing the
waveguide offsets within the parallel architecture can provide a
substantial performance improvement over equal spacing, particularly
when the waveguide height is small. The gain decreases as $d_v/B$
increases and nearly vanishes over an intermediate-height range,
indicating that equal spacing becomes close to the optimized parallel
deployment in this regime. At larger heights, a small gain reappears
because the optimized parallel deployment gradually contracts toward
the centerline, whereas the equally spaced configuration remains
spatially separated. These results show that, even when a parallel
architecture is adopted, optimizing the transverse waveguide offsets
can still provide a noticeable performance benefit.

To illustrate how the optimized multi-waveguide geometry evolves with the
normalized waveguide height, Fig.~\ref{fig:N3_optimal_geometries} shows
three representative deployments for the case of $N=3$ and
$\kappa=A/B=2$. The background depicts the coverage margin relative to
the corresponding worst-location value.
When $d_v/B=0.5$, the optimized deployment is a parallel horizontal
structure, which is consistent with the observation that, at relatively
small waveguide height, the parallel architecture remains highly
effective. When $d_v/B=1.5$, the optimized deployment becomes
nonparallel: one waveguide remains on the horizontal centerline, while
the other two form a symmetric crossing pattern. This shows that, in the
intermediate-height regime, allowing different orientations can improve
the worst-location performance. For $d_v/B=3.0$, all three waveguides numerically collapse
onto the centerline. In this large-height regime, the
benefit of transverse separation becomes weak, and a center-concentrated
deployment is favored.
These representative geometries provide a physical interpretation of the
multi-waveguide optimization results. In particular, they show that the
optimized deployment may evolve from a parallel structure to a
nonparallel mixed-orientation structure, and then to a coincident
centerline structure as the normalized waveguide height increases.

\section{Conclusion}
\label{sec:conclusion} 

This paper investigated long-term waveguide deployment optimization
for pinching-antenna systems over a continuous service region, without
prescribing the waveguide geometry in advance. For the fundamental
two-waveguide case, we characterized the optimal parallel deployment
in closed form and established an explicit geometry-dependent
sufficient condition under which it is also globally optimal. We then
considered the general multi-waveguide case and developed an adaptive
exchange algorithm to jointly optimize the waveguide locations and
orientations. Further analysis showed that parallel waveguides should
be aligned with the longer side of the service region, while in the
large-height regime, the globally optimal unrestricted deployment
asymptotically concentrates on the centerline along the longer side.
Numerical results showed that nonparallel deployments can be
advantageous in intermediate-height regimes, whereas parallel or
center-concentrated structures become favorable in other regimes.
Overall, these results provide analytical guidance for long-term
waveguide deployment in multi-waveguide pinching-antenna systems.

\begin{appendices}
	\section*{\huge Appendices}

\section{Proof of Lemma \ref{lem:parallel_centering}}
\label{appd: lemma vertical coordinate}

First, it is sufficient to consider horizontal waveguide positions
within $[-B,B]$. Indeed, if a waveguide is located above $B$ or below
$-B$, moving it to the nearest boundary weakly decreases its distance
to every service location $y\in[-B,B]$. Hence, an optimal pair can be
written as
\begin{align}
q_1=m-s,\ q_2=m+s,
\end{align}
with $|m|+s\le B$, and therefore $0\le s\le B$.

For a fixed $s$, define
\begin{align}
g_s(u)
=
\frac{1}{d_v^2+(u-s)^2}
+
\frac{1}{d_v^2+(u+s)^2}.
\end{align}
Using $u=y-m$, the worst-location value is
\begin{align}
V_{\rm P}(m,s)
=
\min_{-B-m\le u\le B-m} g_s(u).
\end{align}

We first characterize the shape of $g_s(u)$. The function is even,
and its derivative can be written as
\begin{align}
g_s'(u)
=
\frac{4uH(u^2)}
{[d_v^2+(u-s)^2]^2[d_v^2+(u+s)^2]^2},
\end{align}
where
\begin{align}
H(z)
=
(d_v^2+s^2)(3s^2-d_v^2)
-2(d_v^2+s^2)z-z^2.
\end{align}
Since $H'(z) = -2(d_v^2+s^2)-2z<0, \ z\ge0$, 
$H(z)$ is strictly decreasing. Thus, for $u>0$, $g_s'(u)$ either
remains negative or changes sign at most once from positive to
negative. Therefore, $g_s(u)$ has no nonzero local minimum.
Moreover, since $B\ge s$, both terms in $g_s(u)$ decrease for
$u\ge B$, and hence $g_s(u)$ is decreasing over this range.

We now compare different midpoints. Since $|m|\le B$, the point
$u=0$ belongs to the interval $[-B-m,B-m]$. In addition, one of
its endpoints has absolute value $B+|m|$. Using the evenness of
$g_s(u)$, for any $m\neq0$,
\begin{align}
V_{\rm P}(m,s)
&\le
\min\{g_s(0),g_s(B+|m|)\}
\nonumber\\
&\le
\min\{g_s(0),g_s(B)\}.
\end{align}
For $m=0$, the absence of nonzero local minima implies
\begin{align}
V_{\rm P}(0,s)
=
\min\{g_s(0),g_s(B)\}.
\end{align}
Therefore, we have
\begin{align}
V_{\rm P}(m,s)\le V_{\rm P}(0,s),
\end{align}
showing that $m=0$ is optimal for every $s\in[0,B]$.

Finally, for the resulting centered deployment, the same shape
property of $g_s(y)$ shows that its minimum over $[-B,B]$ can only
occur at the center $y=0$ or at the two boundaries $y=\pm B$.
\hfill$\blacksquare$

\section{Proof of Proposition~\ref{prop:global_parallel}}
\label{app:prop1}

Normalize all distances by $B$ and define
\begin{align}
    \kappa\triangleq\frac{A}{B}\geq1,
    \qquad
    \xi\triangleq\frac{d_v^2}{B^2}.
\label{eq:app_normalized_parameters}
\end{align}
The normalized service region is
$\widetilde{\mathcal A}=[-\kappa,\kappa]\times[-1,1]$.
Consider the axis-aligned ellipse
\begin{align}
    \mathcal E_\kappa
    \triangleq
    \left\{
    (x,y)\in\mathbb R^2:
    \frac{x^2}{\kappa^2}+y^2\leq1
    \right\}
    \subseteq
    \widetilde{\mathcal A}.
\label{eq:app_ellipse}
\end{align}

Consider an arbitrary pair of normalized waveguides
\begin{align}
    \mathcal L_n
    =
    \left\{
    \mathbf z\in\mathbb R^2:
    \mathbf m_n^T\mathbf z=r_n
    \right\},
    \ 
    \|\mathbf m_n\|=1,
    \ n=1,2.
\label{eq:app_general_lines}
\end{align}
Since simultaneously changing the signs of $\mathbf m_n$ and $r_n$
does not change the corresponding line, we choose $r_n\geq0$.
Without loss of generality, let $0\leq r_1\leq r_2$. 
The corresponding normalized received-power sum is given by
\begin{align}
    \widetilde F(\mathbf z)
    \triangleq
    \sum_{n=1}^{2}
    \frac{1}
    {\xi+\left(\mathbf m_n^T\mathbf z-r_n\right)^2}.
\label{eq:app_F}
\end{align}
Since
$\mathcal E_\kappa\subseteq\widetilde{\mathcal A}$, we have 
    $\min_{\mathbf z\in\widetilde{\mathcal A}}
    \widetilde F(\mathbf z)
    \leq
    \min_{\mathbf z\in\mathcal E_\kappa}
    \widetilde F(\mathbf z)$. 
Therefore, to establish the global optimality of the parallel
deployment, it suffices to show that, for any two-waveguide
deployment,  $\min_{\mathbf z\in\mathcal E_\kappa} \widetilde F(\mathbf z)
\leq v_{\rm P}$, where $v_{\rm P}$ is the value achieved by the optimal parallel
deployment.

Under the condition in Proposition~\ref{prop:global_parallel}, we have
\begin{align}
    \xi
    \leq
    \min\left\{
    \frac{5}{4},
    \frac{2\kappa^2-1}{\kappa^2+1}
    \right\}.
\label{eq:app_prop_condition}
\end{align}
In particular, for $\xi\leq5/4$, and hence the first branch of
\eqref{eq:sp_star} applies. The normalized optimal half-separation
is therefore
\begin{align}
    \bar{s}
    \triangleq
    \frac{s_{\rm P}^{\star}}{B}
    =
    \sqrt{\frac{1+\xi}{3}}.
\label{eq:app_sbar}
\end{align}
From \eqref{eq:vp_star}, the corresponding normalized
worst-location value is given by
\begin{align}
    v_{\rm P}
    \triangleq
    B^2V_{\rm P}^{\star}
    =
    \frac{6}{1+4\xi}
    =
    \frac{2}{\xi+\bar{s}^2}.
\label{eq:app_vp}
\end{align}
Meanwhile, the other inequality implied by
\eqref{eq:app_prop_condition} gives
\begin{align}
    \bar{s}^2
    =
    \frac{1+\xi}{3}
    \leq
    \frac{\kappa^2}{\kappa^2+1}.
\label{eq:app_sbar_condition_kappa}
\end{align}

We next upper-bound the worst-location value over
$\mathcal E_\kappa$ for an arbitrary pair of waveguides.
For any unit vector $\mathbf w=[w_x,w_y]^T$, the radial
distance from the origin to the boundary of $\mathcal E_\kappa$
in direction $\mathbf w$ is given by \cite{schneider2014convex}
\begin{align}
    R(\mathbf w)
    =
    \frac{1}
    {\sqrt{\kappa^{-2}w_x^2+w_y^2}}.
\label{eq:app_radial_function}
\end{align}
Thus, $\pm R(\mathbf w)\mathbf w$ lie on the boundary of
$\mathcal E_\kappa$.

Let $\delta\in[0,\pi]$ denote the angle between
$\mathbf m_1$ and $\mathbf m_2$. We first consider
$0\leq\delta\leq\pi/2$. There exist orthonormal vectors
$\mathbf e$ and $\mathbf f$ such that
\begin{align}
    \mathbf m_1
     = \mu\mathbf e+\nu\mathbf f, \ \
    \mathbf m_2
     = \mu\mathbf e-\nu\mathbf f,
\end{align}
where
$\mu=\cos(\delta/2)$ and $\nu=\sin(\delta/2)$.
Let
\begin{align}
    u\triangleq\mu R(\mathbf e),
    \ \
    v\triangleq\nu R(\mathbf f).
\label{eq:app_uv}
\end{align}
Also, for notational convenience, define  $\phi(t)\triangleq\frac{1}{\xi+t^2}$.  
Evaluating $\widetilde F$ at $-R(\mathbf e)\mathbf e$ gives
\begin{align}
    \widetilde F\!\left(-R(\mathbf e)\mathbf e\right)
    &=
    \phi(u+r_1)+\phi(u+r_2)
\nonumber\\
    &\leq
    2\phi(u+r_1),
\label{eq:app_acute_e}
\end{align}
while evaluating it at $R(\mathbf f)\mathbf f$ gives
\begin{align}
    \widetilde F\!\left(R(\mathbf f)\mathbf f\right)
    &=
    \phi(v-r_1)+\phi(v+r_2)
\nonumber\\
    &\leq
    \phi(v-r_1)+\phi(v+r_1).
\label{eq:app_acute_f}
\end{align}

To simplify the subsequent expressions, define
\begin{align}
    G_a(r)
    &\triangleq
    \phi(a-r)+\phi(a+r),
\\
    H_a(r)
    &\triangleq
    2\phi(a+r).
\label{eq:app_GH}
\end{align}
Equations \eqref{eq:app_acute_e} and \eqref{eq:app_acute_f}
then imply
\begin{align}
    \min_{\mathbf z\in\mathcal E_\kappa}
    \widetilde F(\mathbf z)
    \leq
    \min\left\{
    H_u(r_1),G_v(r_1)
    \right\}.
\label{eq:app_acute_bound}
\end{align}

For $\pi/2\leq\delta\leq\pi$, let
$\gamma=(\pi-\delta)/2\in[0,\pi/4]$. There exist orthonormal
vectors $\mathbf e$ and $\mathbf f$ such that
\begin{align}
    \mathbf m_1 = \mu\mathbf e+\nu\mathbf f,
\ \
    \mathbf m_2 = -\mu\mathbf e+\nu\mathbf f,
\end{align}
where $\mu=\cos\gamma$ and $\nu=\sin\gamma$.

Defining $u$ and $v$ as in \eqref{eq:app_uv}, evaluating
$\widetilde F$ at $R(\mathbf e)\mathbf e$ and
$-R(\mathbf f)\mathbf f$ yields
\begin{align}
    \min_{\mathbf z\in\mathcal E_\kappa}
    \widetilde F(\mathbf z)
    \leq
    \min\left\{
    G_u(r_1),H_v(r_1)
    \right\}.
\label{eq:app_obtuse_bound}
\end{align}
It remains to show that the right-hand sides of
\eqref{eq:app_acute_bound} and \eqref{eq:app_obtuse_bound}
cannot exceed $v_{\rm P}$.

To simplify the remaining algebra, let
\begin{align}
    h\triangleq\frac{1}{\kappa^2},
    \ \
    D_e\triangleq R^{-2}(\mathbf e),
    \ \
    D_f\triangleq R^{-2}(\mathbf f).
\label{eq:app_h_DeDf}
\end{align}
Since $\mathbf e$ and $\mathbf f$ are orthonormal, we have
\begin{align}
    D_e+D_f=1+h,
    \ \ 
    h\leq D_e,D_f\leq1.
\label{eq:app_DeDf_properties}
\end{align}
Moreover, \eqref{eq:app_sbar_condition_kappa} becomes
\begin{align}
    \bar{s}^2
    \leq
    \frac{1}{1+h}.
\label{eq:app_sbar_condition}
\end{align}
From \eqref{eq:app_uv}, we have
\begin{align}
    u^2D_e+v^2D_f
    =
    \mu^2+\nu^2
    =
    1.
\label{eq:app_uv_identity}
\end{align}

We first consider \eqref{eq:app_obtuse_bound}. If $v+r_1\geq\bar{s}$, 
then we have
\begin{align}
    H_v(r_1)
    =
    \frac{2}{\xi+(v+r_1)^2}
    \leq
    \frac{2}{\xi+\bar{s}^2}
    =
    v_{\rm P}.
\label{eq:app_H_bound}
\end{align}
It therefore remains to consider $v+r_1<\bar{s}$, which
implies $v<\bar{s}$. From
$D_f=1+h-D_e$ and \eqref{eq:app_uv_identity}, we obtain
\begin{align}
    u^2 &= \frac{1-v^2D_f}{D_e} \nonumber\\
    &= v^2+ \frac{1-(1+h)v^2}{D_e}.
\label{eq:app_u_expression}
\end{align}
Since
$v^2<\bar{s}^2\leq1/(1+h)$ and $D_e\leq1$, we have $u^2 \geq 1-hv^2$. 
Let $\bar u \triangleq \sqrt{1-hv^2}$. 
Then, we have
\begin{align}
    \bar u^2-(\bar{s}-v)^2
    = 1-\bar{s}^2+2\bar{s}v-(1+h)v^2.
\label{eq:app_ubar_difference}
\end{align}
The right-hand side of \eqref{eq:app_ubar_difference} is
concave in $v$ and is nonnegative at both $v=0$ and
$v=\bar{s}$, where \eqref{eq:app_sbar_condition} is used
at $v=\bar{s}$. Hence,
\begin{align}
    \bar u
    \geq
    \bar{s}-v
    >
    r_1.
\label{eq:app_ubar_order}
\end{align}
Since $G_a(r_1)$ decreases with $a$ for $a>r_1$ and
$u\geq\bar u$, we have
\begin{align}
    G_u(r_1)
    \leq
    G_{\bar u}(r_1).
\label{eq:app_Gu_bound}
\end{align}

It therefore remains to establish
\begin{align}
    G_{\bar u}(r)
    \leq
    v_{\rm P},
    \ 
    0\leq r\leq\bar{s}-v.
\label{eq:app_scalar_goal}
\end{align}
Since
$\xi=3\bar{s}^2-1$, clearing the positive denominators in
$v_{\rm P}-G_a(r)$ shows that its sign is determined by
\begin{align}
    \Delta(a,r)
    \triangleq\;&
    a^4-2a^2r^2+2a^2\bar{s}^2-a^2+r^4
\nonumber\\
    &+
    2r^2\bar{s}^2-r^2-3\bar{s}^4+\bar{s}^2.
\label{eq:app_Delta}
\end{align}
For $a^2=1-hv^2$, differentiating $\Delta(a,r)$ with respect
to $r^2$ gives
\begin{align}
    \frac{\partial\Delta(a,r)}{\partial r^2}
    =
    -2a^2+2r^2+2\bar{s}^2-1.
\end{align}
Since $0\leq r\leq\bar{s}-v$, we have
$r^2\leq(\bar{s}-v)^2$. Substituting this inequality together
with $a^2=1-hv^2$ yields
\begin{align}
    \frac{\partial\Delta}{\partial r^2}
    \leq
    -3+4\bar{s}^2-4\bar{s}v+2(1+h)v^2.
\label{eq:app_Delta_derivative_bound}
\end{align}
The right-hand side of \eqref{eq:app_Delta_derivative_bound} is a
convex quadratic function of $v$. Since $0\leq v\leq\bar{s}$, its
maximum over this interval is attained at one of the two endpoints.
At $v=0$, it becomes
\begin{align}
    4\bar{s}^2-3
    \leq 0,
\end{align}
where we have used
$\bar{s}^2=(1+\xi)/3\leq3/4$, which follows from
$\xi\leq5/4$. At $v=\bar{s}$, it becomes
\begin{align}
    -3+2(1+h)\bar{s}^2
    \leq
    -1,
\end{align}
where \eqref{eq:app_sbar_condition} has been used. Therefore,
\begin{align}
    \frac{\partial\Delta(\bar u,r)}{\partial r^2}
    \leq0,
    \ \
    0\leq r\leq\bar{s}-v.
\end{align}
Thus, $\Delta(\bar u,r)$ is nonincreasing in $r^2$, which yields
\begin{align}
    \Delta(\bar u,r)
    \geq
    \Delta(\bar u,\bar{s}-v).
\label{eq:app_Delta_min}
\end{align}

Substituting
$\bar u^2=1-hv^2$ and $r=\bar{s}-v$ into
$\Delta(a,r)$ and simplifying gives
\begin{align}
    \Delta(\bar u,\bar{s}-v)
    =
    v\,P(v),
\label{eq:app_Delta_P}
\end{align}
where
\begin{align}
    P(v)
    \triangleq\;&
    (1+h)^2v^3
    -4(1+h)\bar{s}v^2
\nonumber\\
    &+
    (8\bar{s}^2-3-h)v
    +2\bar{s}(3-4\bar{s}^2).
\label{eq:app_P}
\end{align}
Since $v\geq0$, it remains to show that
\begin{align}
    P(v)\geq0,
    \ 
    0\leq v\leq\bar{s}.
\end{align}

To this end, write $v=\bar{s}z, \ 0\leq z\leq1$, 
and collect the recurring factor $q\triangleq(1+h)\bar{s}^2$. 
From \eqref{eq:app_sbar_condition}, we have $q\leq1$.
Moreover, since $\bar{s}^2\leq3/4$, we have $q \leq \frac{3(1+h)}{4}$. 
Combining these two bounds gives
\begin{align}
    q
    \leq
    \min\left\{
    1,\frac{3(1+h)}{4}
    \right\}.
\label{eq:app_q_bound}
\end{align}

Substituting $v=\bar{s}z$ and
$q=(1+h)\bar{s}^2$ into $P(v)$ gives
\begin{align}
    \frac{1+h}{\bar{s}}P(\bar{s}z)
    =\;&
    q(1+h)^2z^3
    -4q(1+h)z^2
\nonumber\\
    &+
    \left[
    8q-(3+h)(1+h)
    \right]z
\nonumber\\
    &+
    6(1+h)-8q.
\label{eq:app_Pz}
\end{align}
For fixed $z$, the right-hand side of \eqref{eq:app_Pz} is
affine in $q$. Its coefficient with respect to $q$ is given by
\begin{align}
    &(1+h)^2z^3
    -4(1+h)z^2
    +8z-8
\nonumber\\
    &\qquad
    \leq
    4(z-1)(z^2+2)
    \leq0,
\label{eq:app_q_coefficient}
\end{align}
where $0<h\leq1$ and $0\leq z\leq1$ have been used.
Hence, \eqref{eq:app_Pz} is nonincreasing in $q$. It is
therefore sufficient to evaluate it at the largest value of $q$
allowed by \eqref{eq:app_q_bound}.

When $h\geq1/3$, we have
$3(1+h)/4\geq1$, so the largest admissible value is $q=1$.
Substituting $q=1$ into \eqref{eq:app_Pz} yields
\begin{align}
    \frac{1+h}{\bar{s}}P(\bar{s}z)
    =
    (z-1)
    \Big[
    &(1+h)^2z^2
    +(1+h)(h-3)z
\nonumber\\
    &+2-6h
    \Big].
\label{eq:app_case_h_large}
\end{align}
The quadratic function in brackets is convex in $z$. At the two
endpoints,
\begin{align}
    z=0 &: \quad 2-6h\leq0,\\
    z=1 &: \quad 2h(h-3)\leq0,
\end{align}
for $1/3\leq h\leq1$. Hence, the quadratic is nonpositive for
all $z\in[0,1]$. Since $z-1\leq0$,
\eqref{eq:app_case_h_large} is nonnegative.

When $0<h\leq1/3$, we have
$3(1+h)/4\leq1$, so the largest admissible value is
$q=3(1+h)/4$. Substituting this value into
\eqref{eq:app_Pz} gives
\begin{align}
    \frac{1+h}{\bar{s}}P(\bar{s}z)
    =
    \frac{z(1+h)}{4}
    \left[
    3\big((1+h)z-2\big)^2-4h
    \right].
\label{eq:app_case_h_small}
\end{align}
Since $0\leq z\leq1$ and $0<h\leq1/3$, we have
\begin{align}
    2-(1+h)z
    \geq
    1-h
    \geq0.
\end{align}
Therefore,
\begin{align}
    3\big((1+h)z-2\big)^2-4h
    &\geq
    3(1-h)^2-4h
\nonumber\\
    &=
    (1-3h)(3-h)
    \geq0.
\end{align}
Thus, \eqref{eq:app_case_h_small} is also nonnegative.

We have therefore shown that $P(v)\geq0$ for
$0\leq v\leq\bar{s}$. Together with
\eqref{eq:app_Delta_min} and \eqref{eq:app_Delta_P}, this proves
\eqref{eq:app_scalar_goal}. Consequently,
\begin{align}
    \min\left\{
    G_u(r_1),H_v(r_1)
    \right\}
    \leq
    v_{\rm P}.
\label{eq:app_obtuse_final}
\end{align}

The same argument applies to \eqref{eq:app_acute_bound} after
interchanging $u$ and $v$. Specifically, if
$u+r_1\geq\bar{s}$, then $H_u(r_1) \leq v_{\rm P}$. 
Otherwise, $u+r_1<\bar{s}$, and
\eqref{eq:app_uv_identity} gives  $v^2\geq1-hu^2$. 
Repeating the preceding argument with $u$ and $v$ interchanged
then gives $G_v(r_1)\leq v_{\rm P}$. Hence,
\begin{align}
    \min\left\{
    H_u(r_1),G_v(r_1)
    \right\}
    \leq
    v_{\rm P}.
\label{eq:app_acute_final}
\end{align}

Combining \eqref{eq:app_acute_bound},
\eqref{eq:app_obtuse_bound},
\eqref{eq:app_obtuse_final}, and
\eqref{eq:app_acute_final}, we obtain
\begin{align}
    \min_{\mathbf z\in\mathcal E_\kappa}
    \widetilde F(\mathbf z)
    \leq v_{\rm P}
\end{align}
for any two-waveguide deployment. Since
$\mathcal E_\kappa\subseteq\widetilde{\mathcal A}$, this further
implies
\begin{align}
    B^2V_2(\mathcal D_2)
    &=
    \min_{\mathbf z\in\widetilde{\mathcal A}}
    \widetilde F(\mathbf z)
    \leq
    \min_{\mathbf z\in\mathcal E_\kappa}
    \widetilde F(\mathbf z)
    \leq
    v_{\rm P}
    =
    B^2V_{\rm P}^{\star}.
\end{align} 
Therefore,
\begin{align}
    V_2^\star
    \leq
    V_{\rm P}^{\star}.
\end{align}
Since the optimal parallel deployment is itself feasible for the
unrestricted problem, we also have
\begin{align}
    V_2^\star
    \geq
    V_{\rm P}^{\star}.
\end{align}
Combining the two inequalities yields 
$V_2^\star = V_{\rm P}^{\star}$, which completes the proof.
\hfill$\blacksquare$



\section{Proof of Lemma \ref{lem:large_height}}
\label{appd: large height}

For sufficiently large $d_v$, we have the uniform expansion
\begin{align}
    \frac{1}{d_v^2+u^2}
    =
    \frac{1}{d_v^2}
    -
    \frac{u^2}{d_v^4}
    +
    O(d_v^{-6}),
\end{align}
where the remainder is uniform over the compact feasible
deployment and service regions. Applying this expansion to
$F_N(\boldsymbol{\psi};\boldsymbol{\Theta},\mathbf q)$ yields
\begin{align}
    \min_{\boldsymbol{\psi}\in\mathcal{A}}
    F_N(\boldsymbol{\psi};\boldsymbol{\Theta},\mathbf q)
    =
    \frac{N}{d_v^2}
    -
    \frac{Q(\boldsymbol{\Theta},\mathbf q)}{d_v^4}
    +
    O(d_v^{-6}),
\end{align}
where  $Q(\boldsymbol{\Theta},\mathbf q)
    \triangleq
    \max_{\boldsymbol{\psi}\in\mathcal{A}}
    \sum_{n=1}^{N}
    \left(
    \mathbf n^{\mathsf T}(\theta_n)\boldsymbol{\psi}-q_n
    \right)^2$. 

We first justify that the large-height optimal deployment must
approach the minimizer set of $Q(\boldsymbol{\Theta},\mathbf q)$.
Let $Q_{\min} \triangleq \min_{\boldsymbol{\Theta},\mathbf q} Q(\boldsymbol{\Theta},\mathbf q)$, 
and let $(\widehat{\boldsymbol{\Theta}},\widehat{\mathbf q})$ be
any minimizer. For each $d_v$, denote a globally optimal deployment
by
$(\boldsymbol{\Theta}^{\star}(d_v),\mathbf q^{\star}(d_v))$.
By global optimality and the uniformity of the above expansion,
\begin{align}
    Q\!\left(
    \boldsymbol{\Theta}^{\star}(d_v),
    \mathbf q^{\star}(d_v)
    \right)
    \leq
    Q_{\min}
    +
    O(d_v^{-2}).
\end{align}
Since $Q(\boldsymbol{\Theta},\mathbf q)\geq Q_{\min}$, it follows that
\begin{align}
    Q\!\left(
    \boldsymbol{\Theta}^{\star}(d_v),
    \mathbf q^{\star}(d_v)
    \right)
    \rightarrow Q_{\min},
    \ d_v\rightarrow\infty.
\end{align}
The feasible deployment set is compact and $Q$ is continuous.
Therefore, every accumulation point of the globally optimal
deployments must be a minimizer of
$Q(\boldsymbol{\Theta},\mathbf q)$.

It remains to characterize the minimizers of
$Q(\boldsymbol{\Theta},\mathbf q)$. Since $\mathcal{A}$ is centrally
symmetric, both $\boldsymbol{\psi}$ and $-\boldsymbol{\psi}$ belong to
$\mathcal{A}$. Averaging the corresponding quadratic sums gives
\begin{align}
    Q(\boldsymbol{\Theta},\mathbf q)
    \geq
    \max_{\boldsymbol{\psi}\in\mathcal{A}}
    \sum_{n=1}^{N}
    \left(
    \mathbf n^{\mathsf T}(\theta_n)\boldsymbol{\psi}
    \right)^2
    +
    \sum_{n=1}^{N}q_n^2 .
\end{align}
Define
\begin{align}
    \mathbf M
    \triangleq
    \sum_{n=1}^{N}
    \mathbf n(\theta_n)\mathbf n^{\mathsf T}(\theta_n)
    =
    \begin{bmatrix}
        M_{xx} & M_{xy}\\
        M_{xy} & M_{yy}
    \end{bmatrix},
\end{align}
where $M_{xx}+M_{yy}=N$. Then
\begin{align}
    \sum_{n=1}^{N}
    \left(
    \mathbf n^{\mathsf T}(\theta_n)\boldsymbol{\psi}
    \right)^2
    =
    \boldsymbol{\psi}^{\mathsf T}
    \mathbf M
    \boldsymbol{\psi}.
\end{align}
For the rectangular service region
$\mathcal{A}=[-A,A]\times[-B,B]$, we obtain
\begin{align}
    \max_{\boldsymbol{\psi}\in\mathcal{A}}
    \boldsymbol{\psi}^{\mathsf T}\mathbf M\boldsymbol{\psi} =
    NB^2
    \!+\!(A^2\!-\!B^2)M_{xx}
    \!+\!2AB|M_{xy}|.
\end{align}
Hence, we have
\begin{align*}
    Q(\boldsymbol{\Theta},\mathbf q)
    \geq
    NB^2
    +(A^2-B^2)M_{xx}
    +2AB|M_{xy}|
    +
    \sum_{n=1}^{N}q_n^2 .
\end{align*}
Since $A>B$, all terms following $NB^2$ are nonnegative.
Moreover, the deployment with $q_n=0$ and
$\sin\theta_n=0$ for all $n$ achieves
$Q(\boldsymbol{\Theta},\mathbf q)=NB^2$.
Therefore, we have $Q_{\min}=NB^2$, 
and any minimizer must satisfy  $M_{xx}=0, \ q_n=0,\ n=1,\ldots,N$. 
Finally, since $M_{xx} = \sum_{n=1}^{N}\sin^2\theta_n$, 
the condition $M_{xx}=0$ implies
$\sin\theta_n=0$ for every $n$. Thus, every minimizer of
$Q(\boldsymbol{\Theta},\mathbf q)$ consists of coincident
centerline waveguides parallel to the longer side of the service
region.

Consequently, every accumulation point of the globally optimal
deployment as $d_v/B\rightarrow\infty$ has this structure, i.e.,
\begin{align}
    |q_n^{\star}|\rightarrow0,
    \ 
    |\sin\theta_n^{\star}|\rightarrow0,
    \ n=1,\ldots,N,
\end{align}
which completes the proof.
\hfill$\blacksquare$

\end{appendices}


\smaller[1]

\end{document}